\documentclass[aps,prbbib,floatfix,floats]{revtex4}
\usepackage{graphicx,latexsym}
\usepackage{epsfig}
\usepackage{amsmath}
\usepackage{color}
\begin{document}
\title{\bf Analyzing the measured phase in the multichannel Aharonov-Bohm interferometer}
\author{M. \c Tolea$^{1}$, M. Ni\c t\u a$^{1}$, A. Aldea$^{1,2}$}
\affiliation{\hskip-1.4cm $^1$ National Institute of Materials
Physics, POB MG-7, 77125
 Bucharest-Magurele, Romania. \\
$^2$ Institute of Theoretical Physics, Cologne University, 50937
Cologne, Germany.}
\begin{abstract}
We address the quantum dot phase measurement problem in an open
Aharonov-Bohm interferometer, assuming multiple transport channels.
In such a case, the quantum dot is characterized by more than one
intrinsic phase for the electrons transmission. It is shown that the
phase which would be extracted by the usual experimental method
(i.e. by monitoring the shift of the Aharonov-Bohm oscillations, as
in Schuster {\it et al.}, Nature {\bf 385}, 417 (1997)) does not
coincide with any of the dot intrinsic phases, but is a combination
of them.  The formula of the measured phase is given. The particular
case of a quantum dot containing a $S=1/2$ spin is discussed and
variations of the measured phase with less than $\pi$ are found, as
a consequence of the multichannel transport.
\end{abstract}
\maketitle
\section{Introduction}
  The recent developments in the building of mesoscopic devices allow
to measure the phase (phase shift) of the electronic wave function
after it is scattered by a quantum dot. Yacoby {\it et al.}
\cite{Yacoby}, placed a quantum dot in one of the arms of a quantum
ring, realizing an Aharonov-Bohm (AB) interferometer and performed for the first time
a phase measurement. In order to avoid multiple encirclements of the
ring, Schuster {\it et al.} \cite{Schuster} went a step further and opened the Aharonov-Bohm
interferometer by adding a number of additional leads (see also
\cite{Hack, aharony_PE} and  \cite{Aharony,Weidenmuller} about the
conditions of opening the interferometer). In this case, the
electronic waves that reach the drain, have passed through the dot
and through the reference arm only once, with the amplitudes
$t_{dot}=|t_{dot}|e^{i\phi_{dot}}$ and
$t_{ref}=|t_{ref}|e^{i\phi_{ref}}$ respectively, yielding the total
transmittance through the interferometer:
\begin{eqnarray}
T_{int}=|t_{dot}|^2+|t_{ref}|^2+2|t_{dot}||t_{ref}|\cos(\phi_{dot}-\phi_{ref}).
\end{eqnarray}

The dot phase $\phi_{dot}$ is extracted form Eq.1 by varying the
magnetic flux through the interferometer, and in \cite{Schuster} a
phase lapse of $\pi$ was found, between any pair of consecutive
resonances. The expected behavior would have been a phase increase by $\pi$
on each resonance (which concordes with the experimental findings), but also a constant value
 of the phase between resonances (when the dot occupancy does not change), as suggested by the Friedel sum rule \cite{Friedel}. The universal phase lapse behavior found by \cite{Schuster} is one of the longest-standing puzzles in mesoscopic physics, which lead to numerous theoretical
studies -e.g. \cite{Hack,Yeyati,Oreg,K}-  the issue being still
under debate.

In 2005, Avinun-Kalish {\it et al.} \cite{Avinun} launched a couple of new puzzles when performing phase
measurements on a quantum dot that was initially empty and then
gradually filled with electrons. Their experiment showed that the phase lapse of
$\pi$ was actually not universal, as found in \cite{Schuster}, but
appears after the dot is occupied with several electrons. In the
few-electrons regime, the phase variation shows a different
behavior, also intriguing:
 one could notices a number of dips in the
phase evolution of amplitude less than $\pi$, attributed by the
authors to the Kondo effect. However, other experimental papers that
addressed the phase measurement in the Kondo regime (e.g.
\cite{Zaffalon}) reported a different behavior, namely an increase
of the phase by $\pi /2$ on the resonance, which then remains
constant at this value. The paper of Avinun-Kalish {\it et al.}
\cite{Avinun} opened a new trend of calculations on  few-electrons
systems (e.g. \cite{Baksmaty, Karrasch,Bertoni, Gurvitz,Buttiker,
prl}). The reduced variation of the phase on and between (some)
resonances, in the few-electrons regime, is a new puzzle in
mesoscopic physics, and one of the motivations of our paper. We
will show that the existence of multiple transport channels can lead to
a reduced variation of the measured phase.

   To illustrate our statement, we shall address the scattering of an electron
through a quantum dot with a $S=1/2$ magnetic impurity. The model
has the advantage of being exactly solvable, while being a good
approximation for the case when one electron is located in the dot,
and another -itinerant- electron is scattered. Some authors proposed
more complex models for impurity dots, e.g.
\cite{Govorov,Murthy,Kaul,Peeters,Rossier}, but this is not a
purpose of this paper, where a simple model is needed to set the
focus on the phase measurement procedure.

The scattering of an electron on an impurity that possesses internal
degrees of freedom was addressed before in mesoscopic physics, e.g by Mello {\it et al.}
\cite{Mello}, but not in the context of the interferometry and phase
measurement.

The outline of the paper is as follows: in section II we describe the transport
through the multichannel interferometer and derive the formula of
the measured phase; sections III and IV address the case of a
quantum dot with a magnetic impurity; the conclusions are given in
section V.

\section{The measured phase}

We consider an open Aharonov-Bohm interferometer, sketched in Fig.1.
The term "open" is used in the sense of \cite{Schuster,Hack,aharony_PE,Aharony,Weidenmuller}, meaning that multiple
encirclements of the ring are neglected and only one interference is assumed
between the partial electronic waves traveling through the upper and lower arms, respectively.
The upper arm has an embedded quantum dot and the lower arm is the
reference arm. The two arms enclose a magnetic flux $\Phi$. The
interferometer is connected to two transport leads, left (L) and
right (R).

\begin{figure}[ht]
\centering \epsfxsize=0.7\textwidth \epsfbox{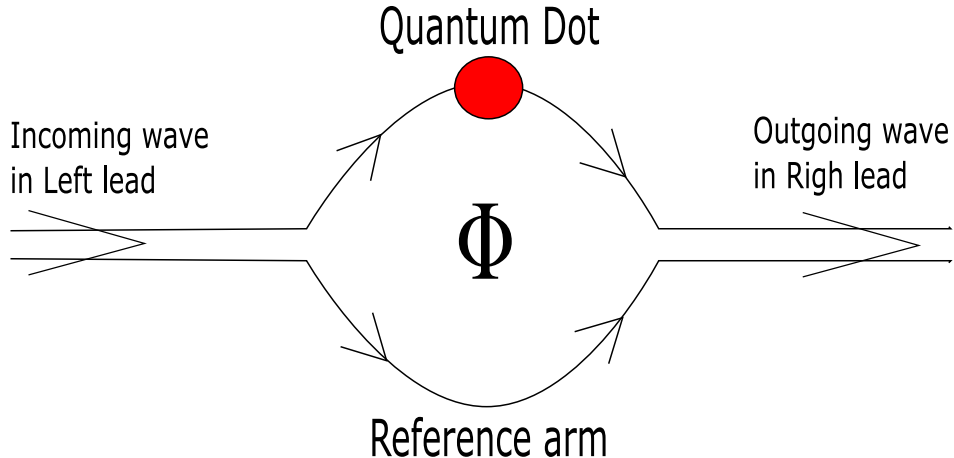} \caption
{(Color online) Scheme of the Aharonov-Bohm intereferometer.}
\label{ab}
\end{figure}

We consider multiple electron transport channels, denoted with $m$.
For every pair of incoming and outgoing channels ($m,m'$), there is
a tunneling process through the dot that has a tunneling amplitude
$t_{m',m}$. The phases $\phi_{m',m}=Arg(t_{m',m})$ are called
intrinsic dot phases. For the tunneling amplitude through the
reference arm, $t_{ref}=|t_{ref}|e^{i\phi_{ref}}$, the phase can be
set equal to the magnetic flux through the ring $\phi_{ref}=\Phi$
(and in the numerical calculations we will take $|t_{ref}|=1$).

We are interested in calculating the total transmittance through the
open interferometer, which is the experimentally available quantity. Incoherent leads are considered, for which the
incident electron comes through each of the channels $m=1,...,M$
with equal probability $1/M$. After a straightforward calculation
(see Appendix A), the total transmittance through the interferometer
can be expressed:
\begin{eqnarray}
T_{int}=T_{dot}+|t_{ref}|^2+\frac2M|t_{ref}|\sum_{m}|t_{m,m}|\cos(\phi_{m,m}-\Phi),
\end{eqnarray}
with $T_{dot}=1/M\sum_{m,m'}|t_{m,m'}|^2$. Now we define the
quantity $t_{phase}$:
\begin{eqnarray}
t_{phase}\equiv \frac1M\sum_{m}t_{m,m},
\end{eqnarray}
and its phase $\phi_{exp}=Arg(t_{phase})$. With these notations, the
total transmittance reads:
\begin{eqnarray}
T_{int}=T_{dot}+|t_{ref}|^2+2|t_{ref}||t_{phase}|\cos(\phi_{exp}-\Phi).
\end{eqnarray}

The above formula resembles Eq.1, but it is no longer the simple
interference of just two waves. In the experiments
\cite{Schuster,Avinun,Zaffalon}, the transmittance through the
interferometer exhibits AB oscillations when $\Phi$ is varied and
from the shift of the oscillations (when a gate is applied on the
dot, for instance) one extracts the evolution of the measured phase.
From Eq.4, it is clear that the measured phase is, in the
multichannel case, $\phi_{exp}$ which can be expressed:
\begin{eqnarray}
\cos{\phi_{exp}}=\frac{\sum_{m}{|t_{m,m}|\cos\phi_{m,m}}}{\sqrt{(\sum_m{|t_{m,m}|\cos\phi_{m,m}})^2+
(\sum_m{|t_{m,m}|\sin\phi_{m,m}})^2}}.
\end{eqnarray}

The above formula is the main formal
result of the paper. When the transport is single-channel, it is obvious that the measured phase $\phi_{exp}$ is equal to the intrinsic dot phase;
for the multichannel transport, the above formula gives the
value of the measured phase $\phi_{exp}$ in terms of the dot
intrinsic phases $\phi_{m,m}$ and amplitude modules $|t_{m,m}|$.

Eq. 5 already allows us to make a comment regarding the evolution of
$\phi_{exp}$, if we assume that $t_{m,m}$ are of resonance
type, with the module presenting a maxima at the resonant energy and
the phase increasing, for instance, from $0$ to $\pi$ on the
resonance width. Then between consecutive (in-phase) resonances, one
has to add two complex numbers that are out of phase. This leads to
a smaller amplitude of $t_{phase}$, but also to a reduced variation
of its phase, in the regions where the resonances overlap. We will
show this explicitly in the example presented in the following.

\section{Quantum dot with a magnetic impurity}

In this section and the next one, we will describe in detail an example of multichannel scattering, with the
aim to obtain in the end the formula of the measurable phase (with the standard experimental method \cite{Schuster,Avinun,Zaffalon}),
for the example considered.

We consider here a
quantum dot containing a $S=1/2$ magnetic impurity , which interacts with
the itinerant electrons via an exchange interaction. First, one must define the scattering channels and calculate the tunneling amplitudes. According to the standard
scattering theory, the impurity dot represents a target that
possesses internal degrees of freedom (the impurity may flip its spin) and, in such a case, the
scattering channels ($|m\rangle$) are the states of the
electron+impurity system:
$|\uparrow\Uparrow\rangle$,$|\downarrow\Downarrow\rangle$,$|\uparrow\Downarrow\rangle$,
and $|\downarrow\Uparrow\rangle$, where the simple arrow stands for
the spin of the itinerant electron and the double arrow stands for
the spin of the impurity (see also \cite{Joshi}, where the same
basis is used to study the spin transmittance through a closed
interferometer).

The Hamiltonian of the dot corresponds to the spin interaction
between the electron and impurity:
\begin{eqnarray}
H^{D}&=& -J\vec{s}\vec{S}=-J(s_zS_Z+\frac12s_+S_-+\frac12s_-S_+)
\end{eqnarray}

The terms in the Hamiltonian ca be written in detail:
\begin{eqnarray}
s_zS_Z&=&\frac14|\uparrow\Uparrow\rangle \langle
\uparrow\Uparrow|+ \frac14|\downarrow\Downarrow\rangle
\langle \downarrow\Downarrow|\nonumber\\
&-&\frac14|\downarrow\Uparrow\rangle \langle
\downarrow\Uparrow|-
\frac14|\uparrow\Downarrow\rangle \langle \uparrow\Downarrow|,\nonumber\\
s_+S_-&=&|\uparrow\Downarrow\rangle \langle \downarrow\Uparrow|,\nonumber\\
s_-S_+&=&|\downarrow\Uparrow\rangle \langle
\uparrow\Downarrow|.
\end{eqnarray}

The quantum dot site index
($n_D$ in Appendix B, where the general case is addressed) can be omitted, since the dot is single-site,
 and there is no risk for confusions (we write for instance
$|\uparrow\Uparrow\rangle$ for the dot site, instead of $|n_{D},\uparrow\Uparrow\rangle$).

\begin{figure}[ht]
\centering
 \epsfxsize=0.7\textwidth \epsfbox{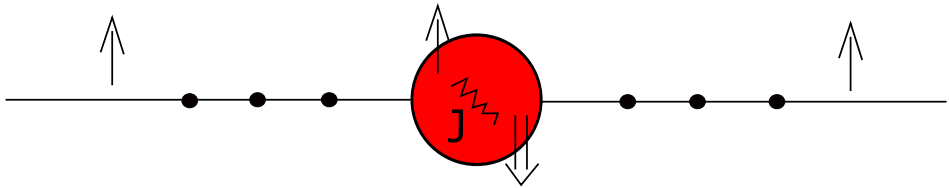}
  \caption {(Color online) Scheme of the quantum dot containing a magnetic impurity, which interacts with
 the itinerant electrons through an exchange interaction $J$. The dot is connected to discrete leads.}
\end{figure}

To calculate the tunneling amplitudes, we connect the dot to
discrete leads, see Fig.2 (for a comparison between discrete and
continuous models in mesoscopic physics, see \cite{Nita}). The
hopping constant between the dot and the leads is $\tau$, and the
hopping between the leads neighboring sites is $\tau_l$ (in the
numerical calculations we will take $\tau_l=1$).

Applying the general recipe in Appendix B (see also \cite{Datta,GA} for equivalent demonstrations) for the particular case of a single site dot
and four transport channels, one writes the effective Hamiltonian (which later shall be used to express the tunneling amplitudes):
\begin{eqnarray} H^D_{eff}=\begin{bmatrix}
-J/4+x & 0   &   0  & 0    \\
0   & -J/4+x &   0  & 0    \\
0   & 0   &  J/4+x & -J/2  \\
0   & 0   &  -J/2 &  J/4+x
\end{bmatrix},
\end{eqnarray}
where $x=2\tau^2 e^{-ik}$ and the columns and rows correspond to the following counting of
channel vectors:
$|\uparrow\Uparrow\rangle$,$|\downarrow\Downarrow\rangle$,$|\uparrow\Downarrow\rangle$
and $|\downarrow\Uparrow\rangle$.

Finally, as shown in Appendix B, the tunneling amplitudes between
the channels $m$ and $m'$ can be expressed as:

\begin{eqnarray}
t_{m',m}=2i\frac{\tau^2}{\tau_l}\sin(k)\langle
m'|\frac{1}{E-H^D_{eff}}|m\rangle,
\end{eqnarray}

There are six
possible processes, corresponding to the non-zero elements of
$1/(E-H^D_{eff})$, and respecting the total spin conservation. To simplify
the index notation, the
non-vanishing tunneling amplitudes will be denoted by $t_i$ with $i=1\cdots 6$:
\begin{eqnarray}
&1)&~|\uparrow\Uparrow\rangle\rightarrow|\uparrow\Uparrow\rangle
;~t_{\uparrow\Uparrow,\uparrow\Uparrow}=t_1=|t_{1}|e^{i\phi_{1}},\nonumber\\
&2)&~|\uparrow\Downarrow\rangle\rightarrow|\uparrow\Downarrow\rangle
;~t_{\uparrow\Downarrow,\uparrow\Downarrow}=t_2=|t_{2}|e^{i\phi_{2}},\nonumber\\
&3)&~|\downarrow\Uparrow\rangle\rightarrow|\uparrow\Downarrow\rangle
;~t_{\uparrow\Downarrow,\downarrow\Uparrow}=t_3=|t_{3}|e^{i\phi_{3}},\\
&4)&~|\downarrow\Uparrow\rangle\rightarrow|\downarrow\Uparrow\rangle
;~t_{\downarrow\Uparrow,\downarrow\Uparrow}=t_4=|t_{4}|e^{i\phi_{4}},\nonumber\\
&5)&~|\uparrow\Downarrow\rangle\rightarrow|\downarrow\Uparrow\rangle
;~t_{\downarrow\Uparrow,\uparrow\Downarrow}=t_5=|t_{5}|e^{i\phi_{5}},\nonumber\\
&6)&~|\downarrow\Downarrow\rangle\rightarrow|\downarrow\Downarrow\rangle
;~t_{\downarrow\Downarrow,\downarrow\Downarrow}=t_6=|t_{6}|e^{i\phi_{6}}.\nonumber
\end{eqnarray}

 From the spin rotation symmetry of the dot Hamiltonian, one has:
\begin{eqnarray}
t_1=t_6,\nonumber \\
t_{2}=t_{4},\\
t_{3}=t_{5}.\nonumber
\end{eqnarray}

During the processes 3
and 5, the spins flip (see  \cite{Konig} for a discussion on the
importance of spin-flip processes for open and closed
interferometers), while the others are non-flip processes.

In Fig.3 we plot the amplitude modulus and phases for the three
tunneling processes $t_1$, $t_2$ and $t_3$ versus the energy of the incident
electron. One notices that $t_1$ is a single resonance process,
while $t_2$ and $t_3$ have two resonances each. The single-resonance
process, $t_1$, has the classical Breit-Wigner structure, and the
phase increases by $\pi$ on the resonance width.

\begin{figure}[ht]
\epsfxsize=1.1\textwidth \epsfbox{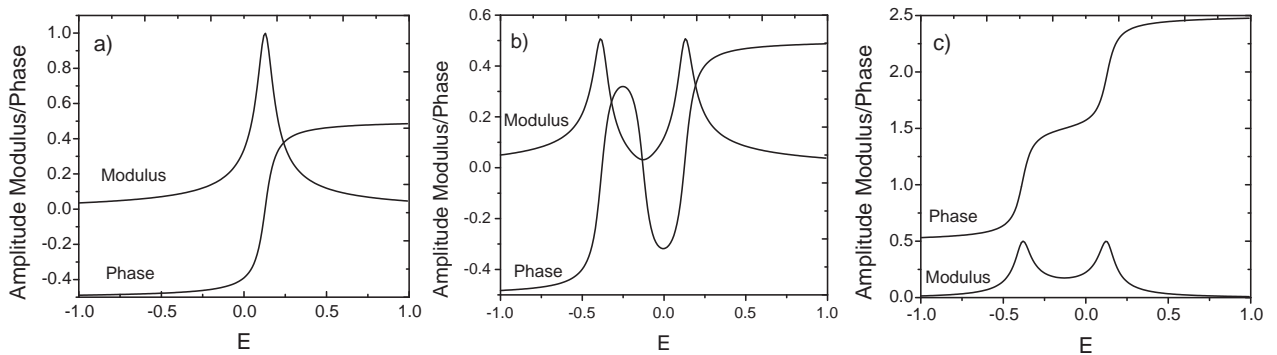} \caption { The modulus
and the phase (in $\pi$ units) of the tunneling amplitude for the
three processes a) $t_1$, b) $t_2$ and c) $t_3$, plotted versus the
energy of the incident electron. The parameters are: $J=-0.5$,
$\tau=0.15$.}
\end{figure}

The processes with two resonances have different structures and they
are of two types: with the two resonances in-phase ($t_2$) or
out-of-phase ($t_3$). For the in-phase case, plotted in Fig.3b
($t_2$), the phase increases from $-\pi/2$ to approx $0.3\pi $ for
the first resonance, then decreases between the resonances to the
value $-0.3\pi$ and finally an increase to $\pi/2$ is noticed on the
second resonance. In this case, the tunneling amplitude between
resonances is lowered, approaching $zero$. For the out-of phase
resonances, Fig.3c ($t_3$), the phase evolution is different, in the
sense that the phase evolves from $\pi/2$ to $3\pi/2$ for the first
resonance, and from $3\pi/2$ to $5\pi /2$ for the second resonance.
Between resonances, the phase is roughly constant and the tunneling
amplitude is increased.

The above discussed behavior of the tunneling amplitudes can be
understood if we perform a spectral decomposition of the dot
Hamiltonian. It is straightforward to show that the dot eigenstates
are: one singlet state with the energy $E_S=3J/4$ and three triplet
states with $E_T=-J/4$. We define $t_S$ as the tunneling amplitude
for the singlet state and  $t_T$ the tunneling amplitude for the
triplet states. One has
$t_S=i\tau^2\sin(k)(\langle\uparrow\Downarrow|-\langle\downarrow\Uparrow|)(E-H^D_{eff})^{-1}
(|\uparrow\Downarrow\rangle-|\downarrow\Uparrow\rangle) =
2i\tau^2\sin(k)/(E-3J/4-2\tau^2 e^{-ik})$. Similarly, one obtains
$t_T=2i\tau^2\sin(k)/(E+J/4-2\tau^2 e^{-ik})$. The tunneling
processes discussed above, $t_{1,2,3}$, can be expressed in the
terms of singlet and triplet (this results straightforwardly from the definition Eq.9; see also
\cite{Aldea}):
\begin{eqnarray}
&t_1&=t_T, \nonumber\\
&t_2&=\frac12(t_T+t_S)~~ in~phase,\\
&t_3&=\frac12(t_T-t_S)~~ out~of~phase.\nonumber
\end{eqnarray}

The sign between the triplet and singlet amplitudes ($t_T$ and
$t_S$) in Eqs.12 explain the in-phase and out-of phase behavior seen
in Fig.3b and c.

Now we can calculate the total transmittance of the dot, $T_{dot}$,
by summing for all incoming channels (the probability of finding the
incoming electron and the impurity in a given spin configuration is
1/4):
\begin{eqnarray}
T_{dot}=\frac14\sum_{i=1}^6|t_{i}|^2=\frac12\sum_{i=1}^3|t_{i}|^2.
\end{eqnarray}

Furthermore, by using  Eqs.12, the dot transmittance becomes
$T_{dot}=1/4|t_S|^2+3/4|t_T|^2$, showing that the singlet resonance
is three times lower in amplitude than the triplet resonance (see
Fig.4a, where for negative $J$ the singlet resonance is lower in
energy), which is an expected result for the two-spin scattering. We
mention that this $1/3$ ratio in the resonances amplitude was also found
by Rejec et. al.\cite{Rejec} for the case of two electrons
scattering and the result was used to give a possible explanation of
the $0.25$ and $0.75$ anomalies in quantum wires.

\section{The measured phase of the magnetic impurity dot}

\begin{figure}[ht]
\centering \epsfxsize=0.65\textwidth \epsfbox{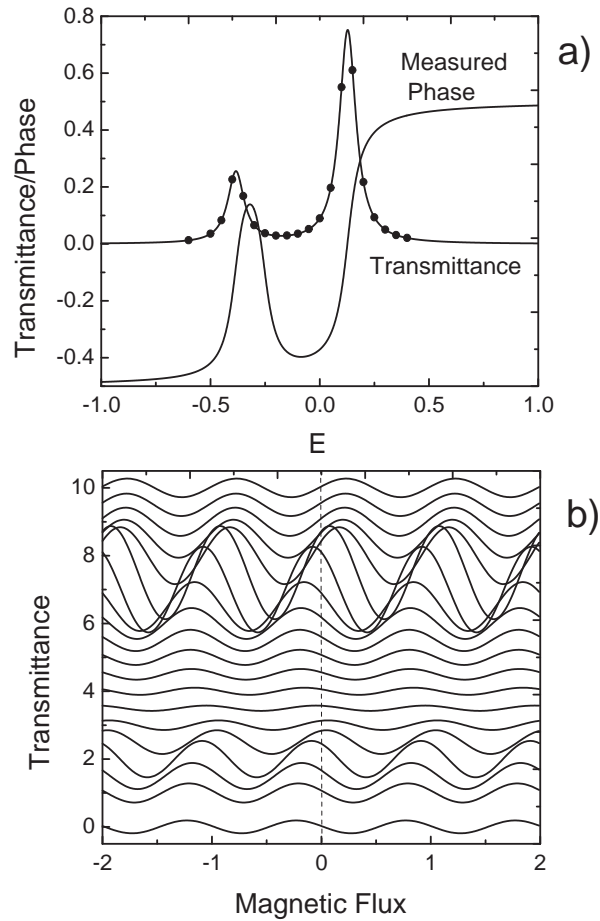} \caption {a)
The transmittance through the quantum dot with a $S=1/2$ magnetic
impurity ($T_{dot}$), and the measured phase ($\phi_{exp}$, given in
in $\pi$ units), plotted versus the energy of the incident electron;
the parameters are: $J=-0.5$, $\tau=0.15$. b) Total transmittance of
the interferometer ($T_{int}$) versus the magnetic flux (in flux
quanta) for the energies $E\in[-0.6,0.4]$ marked with dots in (a)
($E=-0.6$ corresponds to the lowest curve). The curves are shifted
on the vertical for the clarity of presentation. From the horizontal
shift of the Aharonov-Bohm oscillations, one can extract the value
of the measured phase for different energies.}
\end{figure}

Let us now place the impurity dot in one of the arms of an open
Aharonov-Bohm interferometer. The total interferometer transmittance
is the one given by Eq.4, where
$t_{phase}=\frac14(t_1+t_2+t_4+t_6)=\frac12(t_1+t_2)$. Using the
spectral decomposition (Eqs.12), one can write equivalently:
\begin{equation}t_{phase}=\frac14(3\cdot t_T+t_S).\end{equation}

In Fig.4a we plot the quantum dot transmittance $T_{dot}$, given by
Eq.13, and the measured phase $\phi_{exp}=Arg(t_{phase})$ versus the
energy of the incident electron. One can notice that the phase has a reduced
variation (by less than $\pi$) both on the dot resonances and between them.

The Aharonov-Bohm oscillations of the interferometer transmittance
$T_{int}$ are plotted in Fig.4b versus the magnetic flux, in order
to reproduce the experimental method \cite{Schuster,Avinun}. Then,
from the horizontal shift of the Aharonov-Bohm oscillations, one
extracts the quantum dot "measured" phase evolution, plotted in Fig4a.
The dotted vertical line in Fig.4b is
guide for the eye to see the phase evolution. For the lowest curve
(corresponding to $E=-0.6$) one reads a phase $\phi_{exp}$ of about $-\pi /2$ at
zero flux (this means that we have neither a maximum of the
oscillation, nor a minimum, but a value in the middle). The phase
then increases, as the energy of the incoming electron increases,
till the oscillations present a maximum at
$\Phi=zero$ (see the fifth curve from bottom); this means
$\phi_{exp}=0$ (from Eq.4 one sees that, at $zero$ magnetic flux,
$\Phi=0$, the oscillation presents a maximum if, and only if,
$\phi_{exp}=0$). The phase continues to increase a bit more, and
then the AB oscillations begin to shift in the opposite direction,
meaning that the phase starts to decrease, and so on, resulting the
phase evolution in Fig.4a. For the considered parameters, the
measured phase increases with $\sim 0.65\pi$ on the singlet
resonance, then decreases with $\sim 0.55 \pi$ between resonances,
to finally increase again with $\sim 0.9 \pi$ on the triplet peak.
Even if the shape of the phase evolution does not coincide with the
features in \cite{Avinun} - and a more realistic many-body model may
be required - the fact that the measured phase evolves with less
than $\pi$ is recovered by our multi-channel model.

\section{Conclusions}

An experimental method to extract the quantum dot phase, by
monitoring the Aharonov-Bohm oscillations of an open
interferometer, was developed in previous works (e.g.
\cite{Schuster,Avinun,Zaffalon}).

In this paper we show that, when the transport is multichannel, the
phase extracted by using the mentioned method is not an intrinsic
dot phase. Instead, it is the phase of the sum of the diagonal dot
tunneling amplitudes.

We payed a particular attention to the situation when an electron is
scattered by a potential that mixes the transport channels. To this
purpose, we have analyzed the particular case of a quantum dot
containing a $S=1/2$ magnetic impurity. For this simple model, we
have found variations of the measured phase with less than $\pi$,
both on the resonances and between them. This reproduces some
features of a recent experiment \cite{Avinun} and is, in our case, a
consequence of the multi-channel transport.

\section{Acknowledgements}
We acknowledge support from PNCDI II programme under contract No.
515/2009, Contract Nucleu 45N/2009, and of Sonderforschungsbereich
608 at the Institute of Theoretical Physics, University of
Cologne.

\appendix
\section{Total transmittance through a multichannel open interferometer with embedded quantum dot}
The measurable quantity in interference experiments is the total transmittance through the interferometer. To express this, we need the scattering solutions of the Schrodinger
equation. First, we chose the wave function in the left lead to describe
the physical situation when an incident electron is coming from the
left lead in a selected channel $m$ and is reflected back with the
amplitude $r_{m'm}$ into the channel $m'$:
\begin{eqnarray}
\label{a1} |\Psi^{L,m}\rangle=\sum_{n\in
L}e^{ikn}|n,m\rangle+\sum_{n\in L,m'}e^{-ikn}r_{m',m}|n,m'\rangle.
\end{eqnarray}

 The wave function in the right lead describes the
transmitted electron and for the open interferometer it is
considered to be the sum of the two wave functions that correspond
to the two paths: reference and QD arms. The tunneling amplitude
through the reference arm does not depend on the transport channel
$m$, $t_{ref}=|t_{ref}|e^{i\phi_{ref}}$,  and the corresponding wave
function is:
\begin{eqnarray}
|\Psi^{R,m}_{ref}\rangle=t_{ref}\sum_{n\in R}e^{-ikn}|n,m\rangle.
\end{eqnarray}
On the contrary, the tunneling amplitudes for the electron that
passes the quantum dot arm depend on the incoming and outgoing
transport channels $m$ and $m'$, and the wave function reads:
\begin{eqnarray}
|\Psi^{R,m}_{QD}\rangle=\sum_{n\in R, m'} t_{m',m} e^{-ikn}
|n,m'\rangle,
\end{eqnarray}
where $t_{m',m}$ is the transmission amplitude of the quantum dot
for an electron coming in channel m and transmitted in channel m'.

The interference wave function in the right lead of the AB
interferometer will become:
\begin{eqnarray}
|\Psi^{R,m}\rangle && =|\Psi^{R,m}_{ref}\rangle+|\Psi^{R,m}_{QD}\rangle\nonumber\\
       && =\sum_{n\in R}(t_{ref}+t_{m,m})e^{-ikn}|n,m\rangle+\sum_{n\in R,m'\ne m }t_{m',m}e^{-ikn}|n,m'\rangle.
\end{eqnarray}

For the discussed situation, when the incoming channel in the left
lead is $m$, the transmittance of the interferometer is noted by
$T_m$ and is defined as:
\begin{eqnarray}
T_m=|\Psi^{R,m}|^2=|t_{ref}+t_{m,m}|^2+\sum_{m'\ne m}|t_{m',m}|^2.
\end{eqnarray}

Now we consider an incoherent lead for which the incident electron
may come through each of the channels $m$ with equal probability
$1/M$. The total transmittance becomes:
\begin{eqnarray}
T_{int}=\frac 1 M\sum_{m}T_m=\frac 1 M \sum_{m,m'} \big[
|t_{m,m'}|^2 +\delta_{m,m'}|t_{ref}|^2
\nonumber\\+2\delta_{m,m'}|t_{m,m'}||t_{ref}|\cos(\phi_{m,m'}-\phi_{ref})\big].
\end{eqnarray}
or, equivalently,
\begin{eqnarray}
T_{int}=T_{dot}+|t_{ref}|^2+\frac2M|t_{ref}|\sum_{m}|t_{m,m}|\cos(\phi_{m,m}-\Phi),
\end{eqnarray}
where $T_{dot}=1/M\sum_{m,m'}|t_{m,m'}|^2$.

The last term in the interferometer transmittance formula accounts for the quantum interference and depends only on the
 diagonal dot tunneling processes $t_{m,m}$ that do not change the transport channel.
The off-diagonal tunneling amplitudes $t_{m,m'}$ with $m\neq m'$ add
up in module in Eq.A7 and their phases do not influence the total
transmittance (nor can they be measured).

\section{Multi-channel transport through a quantum dot}

This Appendix aims to give an example (for the
completeness of the paper) of how the tunneling amplitudes through a quantum dot can be
calculated in terms of the dot Hamiltonian. See for instance
\cite{Datta,GA} for a similar demonstration using the Green
functions approach. The method presented here calculates the
scattering solution for the system wave function. Although it is a rather general method
(for the case when one electron is scattered by a region with
interaction and internal degrees of freedom), it is practical only for
few-body problems. For many-body problems, the task of calculating
the system wave function is not realistic, and one needs different
approaches.

In this section, we use the same notation as in Appendix A
and consider that the incident electron comes from the left lead
through the channel $m$. The wave functions for the electron in the
left and the right leads were already given in Eqs.A1,A3 and the wave
function for the electron in the dot is:

\begin{eqnarray}
|\Psi^{D,m}_{}\rangle=\sum_{n_D\in {\it D}_{dot},m'}
c_{m',m}(n_D)|n_D,m'\rangle.
\end{eqnarray}

$n_D$ denotes a site in the quantum dot. The coefficients $r$,  $t$ (from Eqs. A1,A3) and $c$ (from Eq. B1) are
parametrically dependent of the incoming channel $m$, and will be
determined form the Schrodinger equation:
\begin{eqnarray}
\begin{bmatrix}
H_D & H_{DL} & H_{DR}\\
H_{LD} & H_{L} & 0 \\
H_{RD} & 0 & H_{R}
\end{bmatrix}
\begin{bmatrix}
|\Psi^{D}\rangle \\
|\Psi^{L}\rangle \\
|\Psi^{R}\rangle
\end{bmatrix}
=E
\begin{bmatrix}
|\Psi^{D}\rangle \\
|\Psi^{L}\rangle \\
|\Psi^{R}\rangle
\end{bmatrix}.
\end{eqnarray}

The leads Hamiltonian correspond to the discrete chain
with the hopping between nearest neighbors taken equal with the energy unit:
\begin{equation}
H_{L(R)}=\sum_{n\in L(R),m'}(|n,m'\rangle \langle n+1,m'|+h.c).
\end{equation}
The coupling Hamiltonians are assumed not to change the channel, and
account for the hopping of the electron between the dot site labeled
$n_D=0_{L(R)}$ and the site of the left (right) lead that is nearest
to the dot and is labeled $1$ (see Fig.2) (like in Eqs.A1-3 the
lead site index does not need to carry the $L$ or $R$- subindex,
there is no danger of confusion):
\begin{eqnarray}
H_{D,L(R)}=\tau_{D,L(R)}\sum_{m'}|0_{L(R)},m'\rangle \langle 1,m'|,\nonumber\\
H_{L(R),D}=\tau_{D,L(R)}^*\sum_{m'}|1,m'\rangle \langle
0_{L(R)},m'|.
\end{eqnarray}

In the following, the coupling coefficients will be taken real and
equal each other $\tau_{DL}=\tau_{DR}=\tau$.

From the solution in the lead, it is straightforward that the
eigenenergy is $E=2\cos(k)$ (this is the dispersion solution for a discrete lead).

In order to calculate the system wave function,  we need to first write
explicitly the set of three equation from  Eq.B2. The steps are
outlined below:

1. One of the equations reads: $H_{LD}\Psi^{D}+H_L\Psi^{L}=E\Psi^L$.
We use the explicit expression of wave function $\Psi^L$ (Eq.A1) and
write this equation for the site 1 of the left lead:
\begin{eqnarray}
(e^{2ik}|1,m\rangle+\sum_{m'}r_{m',m}e^{-2ik}|1,m'\rangle)+\tau\sum_{m'}c_{m',m}(0_L)|1,m\rangle=\nonumber\\
E(e^{ik}|1,m\rangle+\sum_{m'}r_{m',m}e^{-ik}|1,m'\rangle).
\end{eqnarray}
Using Eq.A3 for the right lead we have:
\begin{equation}
\sum_{m'}t_{m',m}e^{-2ik}|1
m'\rangle+\tau\sum_{m'}c_{m',m}(0_R)|1,m'\rangle=E\sum_{m'}t_{m',m}e^{-ik}|1,m'\rangle.
\end{equation}

One multiplies with $\langle 1,m'|$ on the left of Eqs. B5 and B6,
and we have a relation between the dot wave function coefficients
$c_{m',m}$ and the lead wave function coefficients, $r_{m'm}$ and
$t_{m'm}$:

\begin{eqnarray}
&&t_{m',m}=\tau c_{m',m}(0_R),\nonumber\\
&&r_{m',m}=\tau c_{m',m}(0_L)-\delta_{m,m'}.
\end{eqnarray}

2. The equation for the dot function reads:
$H_D\Psi^D+\tau_{LD}\Psi^L_1+\tau_{RD}\Psi^R_1=E\Psi^D$. We
introduce the explicit form of $\Psi^L$ and $\Psi^R$ from Eq.A1,A3 and
we have:
\begin{equation}H_D|\Psi^{D}_{m}\rangle+\tau(e^{ik}|0_L,m\rangle+\sum_{m'}
r_{m',m}e^{-ik}|0_L,m'\rangle)+\tau
\sum_{m'}t_{m',m}e^{-ik}|0_R,m'\rangle=E|\Psi^{D}_{m}\rangle.
\end{equation}

Using the relation between the coefficients of the lead wave
functions and dot wave function from Eqs.B7 one obtains (after some
straightforward algebra) an equation for the coefficients
$c_{m',m}$, which can be written in the compact form:
\begin{equation}(E-H^{D}_{eff})|\Psi^{D}_{m}\rangle=2i\tau (sink)|0_L,m\rangle,\end{equation}
with
\begin{equation}H^{D}_{eff}=H^{D}+\tau^2 e^{-ik}\sum_{m',\alpha=L,R} |0_\alpha,m'\rangle\langle 0_\alpha,m'|\end{equation}

3. Finally, the quantity of interest,
the channel-dependent transmission coefficient, is:

\begin{equation}t_{m',m}=2i\tau^2\sin(k)\langle 0_R,m'|\frac{1}{E-H^D_{eff}}|0_L,m\rangle .
\end{equation}

\end{document}